# Experimental entanglement of six photons in graph states

Chao-Yang Lu[1], Xiao-Qi Zhou[1], Otfried Gühne[2], Wei-Bo Gao[1], Jin Zhang[1], Zhen-Sheng Yuan[1], Alexander Goebel[3], Tao Yang[1], and Jian-Wei Pan[1,3]

1) Hefei National Laboratory for Physical Sciences at Microscale and Department of Modern Physics, University of Science and Technology of China, Hefei, Anhui 230026, People's Republic of China.

2) Institut für Quantenoptik und Quanteninformation, Österreichische Akademie der Wissenschaften, Technikerstraβe 21A, A-6020 Innsbruck, Austria.

3) Physikalisches Institut, Universität Heidelberg, Philosophenweg 12, D-69120 Heidelberg, Germany

**Graph states[1-3] are special kinds of multipartite entangled states that correspond to mathematical graphs where the vertices take the role of quantum spin systems and the edges represent interactions. They not only provide an efficient model to study multiparticle entanglement[1], but also find wide applications in quantum error correction[3], multi-party quantum communication[4] and most prominently, serve as the central resource in one-way quantum computation[5]. Here we report the creation of two special instances of graph states, the six-photon Greenberger-Horne-Zeilinger states[6] -- the largest photonic Schrödinger cat, and the six-photon cluster states[2] -- a state-of-the-art one-way quantum computer. Flexibly, slight modifications of our method allow creation of many other graph states. Thus we have demonstrated the ability of entangling six photons and engineering multiqubit graph states, and created a test-bed for investigations of one-way quantum computation[7-15] and studies of multiparticle entanglement as well as foundational issues such as nonlocality[16,17] and decoherence[18].**

Entanglement lies at the heart of quantum mechanics and plays a crucial role in quantum information processing. Many efforts have been undertaken to create especially multipartite entangled states in different physical systems[19-22]. In recent years, special kinds of multipartite entangled states, the graph states[1-3], have moved into the center of attention. They can be associated with graphs where each vertex represents a qubit prepared in the state $\frac{1}{\sqrt{2}}(|0\rangle+|1\rangle)$ and each edge represents a controlled phase gate having been applied between the two connected qubits. An interesting feature is that many entanglement properties of graph states are closely related to their underlying graphs. Besides their thought-provoking theoretical structure[1], the graph states also have shed new insights in studies of nonlocality[16,17] and decoherence[18] and served as essential resource for various quantum information tasks[3,4], most prominently as the exceptionally universal resource for one-way quantum computation[5]. Encouraging progresses[7-14,20] have been achieved in this direction, especially in the linear optics regime[15]. Yet a major challenge ahead lies in the experimental generation of multiqubit graph states.

Of special interest in the graph state family are the Greenberger-Horne-Zeilinger (GHZ) states and the cluster states. Experimentally, six-atom GHZ states[21] and four-photon cluster states[20] have been realized. Here we report the creation of six-photon GHZ states and cluster state with verifiable six-partite entanglement. To do so, we start from three Einstein-Podolsky-Rosen (EPR) entangled photon pairs in the state

$$|\Phi^+\rangle_{ij} = \frac{1}{\sqrt{2}}(|H\rangle_i|H\rangle_j+|V\rangle_i|V\rangle_j),$$

where *H* and *V* denote horizontal and vertical polarization, and *i* and *j* label the spatial modes of the photons (Fig. 1a). We superpose photons in mode 2 and 3 (4 and 5) at polarizing beam splitters (PBS). Since the PBS transmits *H* and reflects *V* polarization, only if both incoming photons have the same polarization can they go to different outputs. Thus a coincidence detection of all the six outputs corresponds to the state

$$|G_6\rangle = \frac{1}{\sqrt{2}}(|H\rangle_1|H\rangle_2|H\rangle_3|H\rangle_4|H\rangle_5|H\rangle_6 + |V\rangle_1|V\rangle_2|V\rangle_3|V\rangle_4|V\rangle_5|V\rangle_6),$$

which is a six-photon Greenberger-Horne-Zeilinger (GHZ) state, exhibiting an equal superposition of two maximally different quantum states.

By applying a Hadamard (H) gate on photon 4 before it enters into PBS (Fig. 1a), the above scheme can be readily modified to generate a six-photon cluster state. Consider it in two steps: *1*. combine photon 2 and 3, based on a coincidence detection we get a four-photon GHZ state $\frac{1}{\sqrt{2}}(|H\rangle_1|H\rangle_2|H\rangle_3|+\rangle_4+|V\rangle_1|V\rangle_2|V\rangle_3|-\rangle_4)$, where $|\pm\rangle=\frac{1}{\sqrt{2}}(|H\rangle\pm|V\rangle)$; *2*. combine photon 4 and 5, and by a similar reasoning we obtain what we call here a six-photon cluster state

$$|C_6\rangle = \frac{1}{2}(|H\rangle_1|H\rangle_2|H\rangle_3|H\rangle_4|H\rangle_5|H\rangle_6+|H\rangle_1|H\rangle_2|H\rangle_3|V\rangle_4|V\rangle_5|V\rangle_6 \\ +|V\rangle_1|V\rangle_2|V\rangle_3|H\rangle_4|H\rangle_5|H\rangle_6-|V\rangle_1|V\rangle_2|V\rangle_3|V\rangle_4|V\rangle_5|V\rangle_6).$$

For an intuitive understanding, we show in Fig. 1 the underlying graph of the above states and how they grow from smaller (two-qubit) graph states. Up to local unitary transformations, the GHZ states correspond to star-shaped graph, and the cluster state to lattice graph. The effect of combining two photons at PBS can be described by the operator $|HH\rangle\langle HH|+|VV\rangle\langle VV|$, leading to fuse two separate graph states into a single one[8,10]. Specifically, Fig. 1c(d) shows when a two-qubit graph state is combined with the root (leaf) node of a four-qubit star graph state, a six-qubit GHZ (cluster) state is produced.

A nice feature of the graph state representation is that many properties of graph states and their potential use in quantum information processing could be revealed by their underlying graph. For example, the star-graph states have multiple leaf nodes, which is referred as microcluster in Ref. 7,12,13 and can be used in the so-called parallel fusion for building up large cluster states. The graph of the six-qubit cluster state forms a standard quantum circuit under the one-way computer model[20]. Moreover, its geometry embodies a tree-shaped graph which is the basic building block for loss-tolerant one-way quantum computing[14]. Another interesting feature of the cluster state is that not only itself, but even the remaining mixed four-qubit state after two qubits being traced out leads to an all-versus-nothing argument for nonlocality[16], showing a surprisingly strong entanglement persistency.
3

Let us now proceed with the experimental demonstration. In our experiment, we use spontaneous down conversion (SPDC) to produce entangled photons[23]. We make various efforts to prepare high-brightness and stable sources of entangled photons. The setup is illustrated in Fig. 2. A pulsed ultraviolet (UV) laser successively passes through three $\beta$-barium borate (BBO) crystals to generate entangled photon pairs in spatial modes 1-2, 3-4 and 5-6. The photon pairs are prepared in the state $|\Phi^+\rangle$ with an average twofold coincidence count of about $9.3 \times 10^4 s^{-1}$ and a visibility of 93% (91%) in the H/V (+/-) basis. We then superpose the photons 2(4) and 3(5) at the PBS. To achieve good spatial and temporal overlap, the photons are spectrally filtered ($\Delta\lambda_{FWHW} = 3.2 nm$) and detected by fiber-coupled single-photon detectors. By making fine adjustments of delay $\Delta d_1$ ($\Delta d_2$) we are able to observe interference fringes of four-photon entanglement with a visibility of 73% (71%) in mode 1-2-3-4 (3-4-5-6), indicating that the fusion operations have been successfully implemented (see the Supplementary material).

Now we analyze the experimental data of six-photon graph states and characterize the entanglement produced here. Let us first discuss to which extent the desired states were produced and the presence of genuine multipartite entanglement. The quality of the states can be judged by the fidelity, that is, the overlap of the produced state with the desired one. The notion of genuine multipartite entanglement characterizes whether generation of the state requires interaction of all parties: a pure state $|\Psi\rangle$ is called biseparable, whenever a grouping of the six parties into two groups $G_A$ and $G_B$ can be found, such that the state is a product state, that is $|\Psi\rangle = |\alpha\rangle_{G_A} \otimes |\beta\rangle_{G_B}$, otherwise it is genuine multipartite entangled. Consequently, a mixed state is called biseparable, if it is a mixture of biseparable pure states, otherwise it is genuine multipartite entangled.

In order to prove multipartite entanglement, we use the method of entanglement witnesses[24]. An entanglement witness is an observable which has a positive expectation value on all biseparable states. Thus a negative expectation value proves the presence of genuine multipartite entanglement. In what follows we derive efficient entanglement witnesses that are both robust against realistic noise and economical for experimental efforts.

For the six-photon GHZ state we use witness

$$W_G = \frac{I}{2} - |G_6\rangle\langle G_6|.$$

We decompose $|G_6\rangle\langle G_6|$ into locally measurable observables

$$|G_6\rangle\langle G_6| = \frac{1}{2}[(|H\rangle\langle H|)^{\otimes 6} + (|V\rangle\langle V|)^{\otimes 6}] + \frac{1}{12}\sum_{n=-2}^{3}(-1)^n M_{(n)}^{\otimes 6},$$

where $M_{(n)} = \cos(n\pi/6)\sigma_x + \sin(n\pi/6)\sigma_y$ are measurements in the *x-y* plane. To implement this witness, seven measurement settings are required. Fig. 3 shows the measurement results, yielding $Tr(W_G \rho_{exp}) = -0.093 \pm 0.025$, which is negative by 3.7 standard derivations and thus proving the presence of genuine six-partite entanglement.

From the expectation value of the witness, we can directly determine the obtained fidelity as $F_{G_6} = \langle G_6|\rho_{exp}|G_6\rangle = 0.593 \pm 0.025$, which is a considerable improvement of the fidelity of the six-atom GHZ states[21] ( $F = 0.509 \pm 0.004$ ).

For the cluster state a possible witness would be $W_C = I/2 - |C_6\rangle\langle C_6|$. Similar to the constructions of Ref. 25 we use a slightly different witness $\widetilde{W}_C$ which implementation requires only six measurements (see Methods). Fig. 4 shows the measurement results in basis $\sigma_z^{\otimes 3}\sigma_x^{\otimes 3}$ ( $\sigma_x^{\otimes 3}\sigma_z^{\otimes 3}$ ), which together with those of the four other bases $\sigma_z^{\otimes 3}M_{(\pm 1)}^{\otimes 3}$ and $M_{(\pm 1)}^{\otimes 3}\sigma_z^{\otimes 3}$ (see the Supplementary material), gives $Tr(\widetilde{W}_C \rho_{exp}) = -0.095 \pm 0.036$. Thus the genuine six-partite entanglement of the cluster state is also proved. Furthermore, from this result we can obtain a lower bound of the fidelity of our cluster state as $F_{C_6} \geq 0.595 \pm 0.036$.

For an investigation of the bipartite entanglement properties of these graph states we estimate the entanglement of formation from the expectation value of the witness[26]. Here different bipartitions arise when the six parties are divided into two groups. The entanglement of formation $E_F(\rho)$ is an entanglement measure for bipartite systems, quantifying how many EPR pairs are needed for the formation of the state[27]. For the GHZ state we find that for all bipartitions at least $E_F(\rho_{exp}) \geq 0.073 \pm 0.032$. For the cluster state $E_F(\rho_{exp})$ is also always positive, for some bipartitions it is even $E_F(\rho_{exp}) \geq 0.729 \pm 0.106$. A full discussion, also for a different entanglement measure, is given in the Supplementary material.

The imperfections of our graph states are mainly caused by two reasons. First, high-order emissions of entangled photons give rise to the undesired components in *H/V* basis (Fig. 3a). Second, the partial distinguishability of independent photons causes some incoherent mixtures. In spite of the imperfections, genuine entanglement of the six-photon graph states are strictly confirmed. It is possible to improve the fidelity in future experiments, e.g., by using photon-number discriminating detectors to filter out the events of double emissions of photon pairs. Moreover, graph states with high purity can be obtained efficiently using the existing entanglement purification scheme[28]. The linear optical elements such as the PBS may offer a high-accuracy tool for this task[29]. It leaves a crucial open question of how to reach the noise thresholds for optical cluster-state quantum computation[13].

Some further remarks are deserved here. We generate the graph states conditioned on that there is one and only one photon in each of the six outputs. This post-selective feature, on the one hand, together with the fusion method provides a flexible way to create various multiphoton graph states. Slight modifications of our experimental setup will readily allow creation of many other graph states, e.g., six-qubit linear and Y-shaped graph states (see the Supplementary Information). Such a capacity creates a useful multiqubit graph state test-bed. On the other hand, this feature does not prohibit subsequent applications such as tests of quantum nonlocality[16,17] and in-principle verifications of linear optical QIP tasks where photons need to eventually detected[15]. Finally, one may concern about the scalability problem.

Here we refer to Ref. 10 for a recent scheme of scalable tree-graph state generation, which has shown that this obstacle could be overcome.

In conclusion, we have realized two special graph states, the six-photon GHZ state-- the largest photonic Schrödinger cat, and the six-photon cluster state-- a state-of-the-art one-way quantum computer. We have demonstrated the ability of entangling six photons and engineering multiqubit graph state, and creates a versatile test-bed for implementations of sophisticated quantum algorithms[5,11], demonstrations of basis elements of loss- and fault-tolerant one-way quantum computation[12-15], as well as studies of multiparticle entanglement and foundational issues such as non-locality[16,17] and decoherence[18]. The high efficiency and flexibility of the six-photon graph state generation we demonstrated here suggest that the photons manipulated with linear optics are extremely promising candidates for engineering of multiqubit graph states. Our experimental methods incorporate key modules of the efficient graph-state construction schemes[8-10] and fault-tolerant cluster state quantum computer architectures[12-15], thus have profound implications in the burgeoning field of one-way quantum computation. Furthermore, our experiment also enables immediate applications such as quantum error-correction encoding[3] and multiparticle entanglement swapping[30]. Lastly, the efficient entanglement detection method described here will find further applications in future studies of multiparticle entanglement with higher number of qubits.

**Methods:**

The witness for the cluster state can be constructed as follows. Using the results of Ref. 25, the observable $W_C = I/2 - |C_6\rangle\langle C_6|$ is a witness detecting genuine multipartite entanglement around the cluster state. Then we consider the observable

$$\widetilde{W}_C = \frac{3}{2}I - \prod_{i=1,3,5}\frac{g_i+I}{2} - \prod_{i=2,4,6}\frac{g_i+I}{2} - \frac{1}{2}(I \otimes A_0 + A_0 \otimes I) - (A_1 \otimes B_1 + B_1 \otimes A_1)$$
$$= \frac{I}{2} - |C_6\rangle\langle C_6| + |\widetilde{C}_6\rangle\langle \widetilde{C}_6|$$

where the $g_i$ denote the stabilizing operators of the cluster state (see Fig. 4A). Further, we use $A_0 = I - |HHH\rangle\langle HHH| - |VVV\rangle\langle VVV|$, $A_1 = |VVV\rangle\langle VVV| - |HHH\rangle\langle HHH|$, and $B_1 = (M_{(1)}^{\otimes 3} + M_{(-1)}^{\otimes 3})/2\sqrt{3}$, where $M_{(i)}$ defined as for the GHZ state. Finally, $|\widetilde{C}_6\rangle$ denotes a cluster state with different signs, namely

$$|\widetilde{C}_6\rangle = (-|HHHHHH\rangle + |HHHVVV\rangle + |VVVHHH\rangle + |VVVVVV\rangle)/2$$

It is clear that $\widetilde{W}_C - W_C \geq 0$ which implies that $\widetilde{W}_C$ is a valid witness[25]. Further, this implies that the fidelity of the cluster state can be estimated as $F_{C_6} = \langle C_6 | \rho_{\exp} | C_6 \rangle \geq \frac{1}{2} - \langle \widetilde{W}_C \rangle$.

The witness $\widetilde{W}_C$ detects genuine entanglement from the states of the form $\rho(p) = p|C_6\rangle\langle C_6| + (1-p)I/64$ for $p > 0.5$. The determination of the expectation value of the witness $\widetilde{W}_C$ requires six measurement settings, namely and $\sigma_z^{\otimes 3}\sigma_x^{\otimes 3}$, $\sigma_x^{\otimes 3}\sigma_z^{\otimes 3}$, $\sigma_z^{\otimes 3} M_{(\pm 1)}^{\otimes 3}$ and $M_{(\pm 1)}^{\otimes 3}\sigma_z^{\otimes 3}$. The results are shown in Fig. 4 in the main text and in Fig. 3 of the supplementary material.

**References:**

1. Hein, M., Eisert, J. & Briegel, H. J., Multiparty entanglement in graph states. *Phys. Rev. A* **69**, 062311 (2004).

2. Briegel, H. J. & Raussendorf, R., Persistent entanglement in arrays of interacting particles. *Phys. Rev. Lett.* **86**, 910 (2001).

3. Schlingemann, D., & Werner, R. F., Quantum error-correcting codes associated with graphs. *Phys. Rev. A* **65**, 012308 (2002).

4. Clever, R., Gottesman, D. & Lo, H.-K., How to share a quantum secret. *Phys. Rev. Lett.* **83**, 648 (1999).

5. Raussendorf, R. & Briegel, H. J., A one-way quantum computer. *Phys. Rev. Lett.* **86**, 5188 (2001).

**Acknowledgments**: We thank H. J. Briegel, T. Rudolph, D. Browne and S. Yu for helpful discussions. This work was supported by the National Natural Science Foundation of China, the Chinese Academy of Sciences. This work was also supported by the Alexander von Humboldt Foundation, the Marie Curie Excellence Grant of the EU, the FWF, the DFG and the EU(Scala, Olaqui, Prosecco, Quprodis) .

**Competing Interests**

The authors declare that they have no competing financial interests.

Correspondence and requests for materials should be addressed to

C.-Y. Lu (cylu@mail.ustc.edu.cn) or J.-W. Pan (jian-wei.pan@physi.uni-heidelberg.de).

**Figure Captions**

**Figure 1 a)**, Scheme to generate six-photon GHZ states and cluster states by combining three pairs of entangled photons at PBS. The Hadamard (H) gate is inserted for generation of the six-photon cluster states. **b)-d)**, Underlying graph of the six-photon graph states and how they are created by fusion operations. The graph state can thought of as constructed by first preparing the qubits at each vertex in the state $|+\rangle = \frac{1}{\sqrt{2}}(|H\rangle+|V\rangle)$ and then applying controlled phase gates between pairs of neighboring qubits. Here we use the fusion operations, that is, combining photons at PBS, to generate multiqubit graph state efficiently. In the star graph, we refer to the central node as a root and the others as leaves.

**Figure 2** Experimental setup for generation of six-photon graph states. Pumped by a continuous-wave (CW) green laser, the Mode-locked Ti:Sapphire laser outputs a pulsed infrared (IR) laser with a central wavelength of 788 nm, a pulse duration of 120 fs and a repetition rate of 76 MHz, which passes through a LBO ($LiB_3O_5$) crystal (mounted on a motorized translation stage) and up-converted to UV laser ($\lambda = 394 nm$). The UV laser is circulated and focused on the three BBO crystals to produce three pairs of entangled photons. The entangled photons are spectrally filtered by narrow-band filters (with peak transmission rate ~ 98%) and then detected by fiber-coupled single photon detectors ($D_{1T}$, ... $D_{6R}$). We use a programmable multi-channel coincidence unit to register the multifold coincidence events. For polarization analysis, half- and quarter-wave plates (HWP, QWP) together with polarizers or PBS are used. Simply by changing the angle ($\theta$) of the HWP at path 4, our setup is tunable to generate the six-photon GHZ states ($\theta = 0°$) and cluster state ($\theta = 22.5°$).

**Figure 3** Experimental result of the six-photon GHZ state. **a)**, Sixfold coincidence counts in *H/V* basis in 3 hours. **b)**, The expectation values of $M_{(n)}^{\otimes 6}$, each derived from a complete set of 64 sixfold coincidence events in 2 hours in measurement basis $|H\rangle \pm e^{in\pi/6}|V\rangle$.

**Figure 4 a)**, The graph corresponds to the cluster state $|C_6\rangle$ under H transformations on qubit 1, 3, 4, 6, and its stabilizer operators $g_i$, where *i* labels the qubits and $X = \sigma_x, Y = \sigma_y, Z = \sigma_z$. The graph state is a common eigenstate of these stabilizer operators, that is, $g_i|C_6\rangle = |C_6\rangle$, which describe the correlations in the state. The cluster state is the unique state fulfiling this, which allows for an alternative definition of it. **b)-c)**, Sixfold coincidence counts measured in the $\sigma_z^{\otimes 3}\sigma_x^{\otimes 3}$ and $\sigma_x^{\otimes 3}\sigma_z^{\otimes 3}$ basis in 3 hours, from which the expectation values of stabilizer operators $g_{1,3,5}$ and $g_{2,4,6}$ are derived.

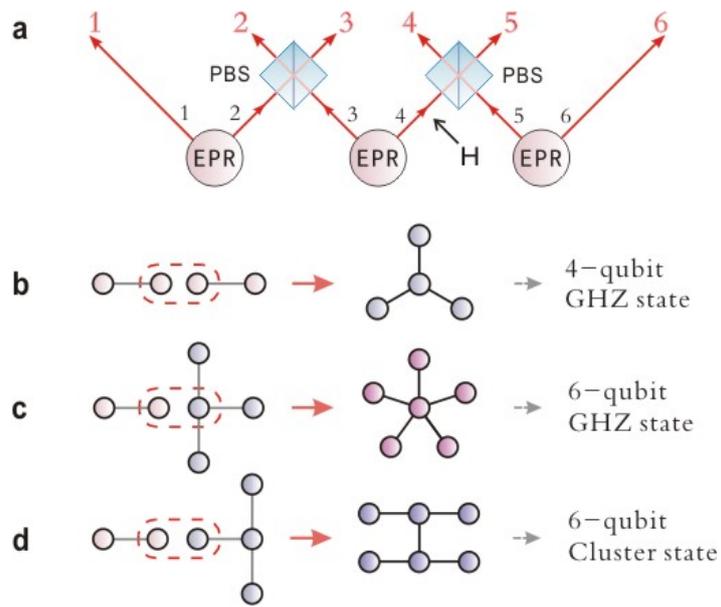

Figure 1

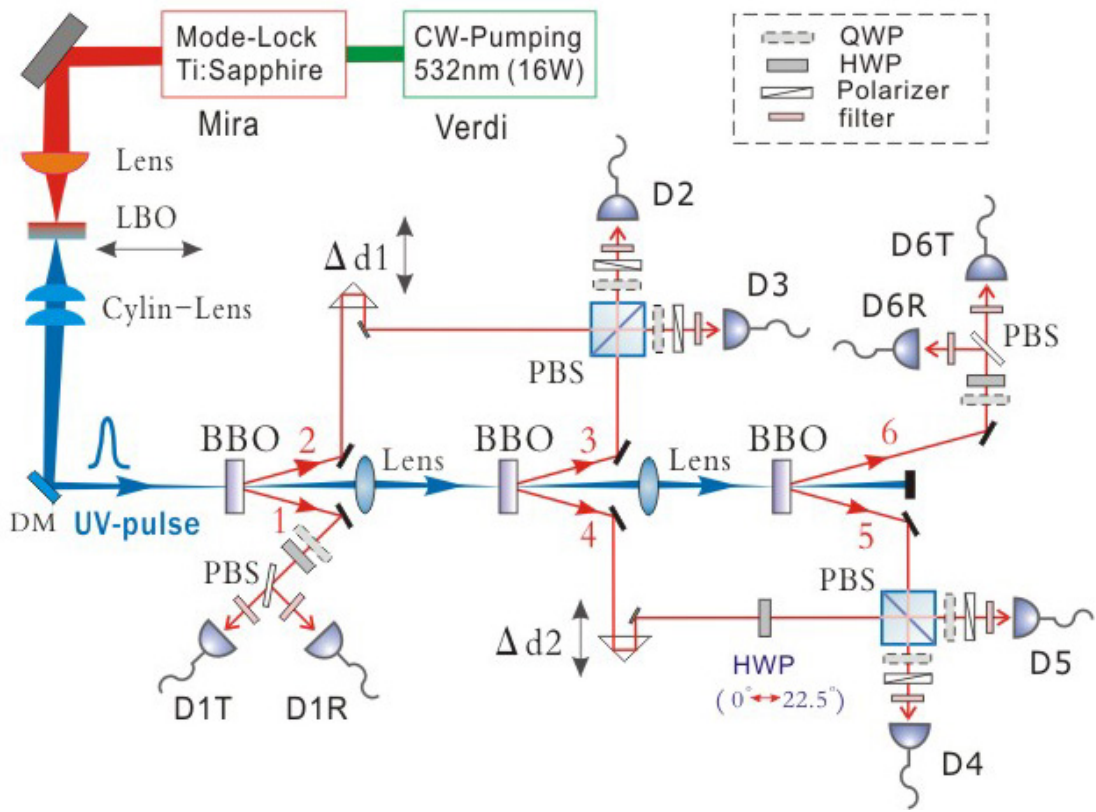

Figure 2

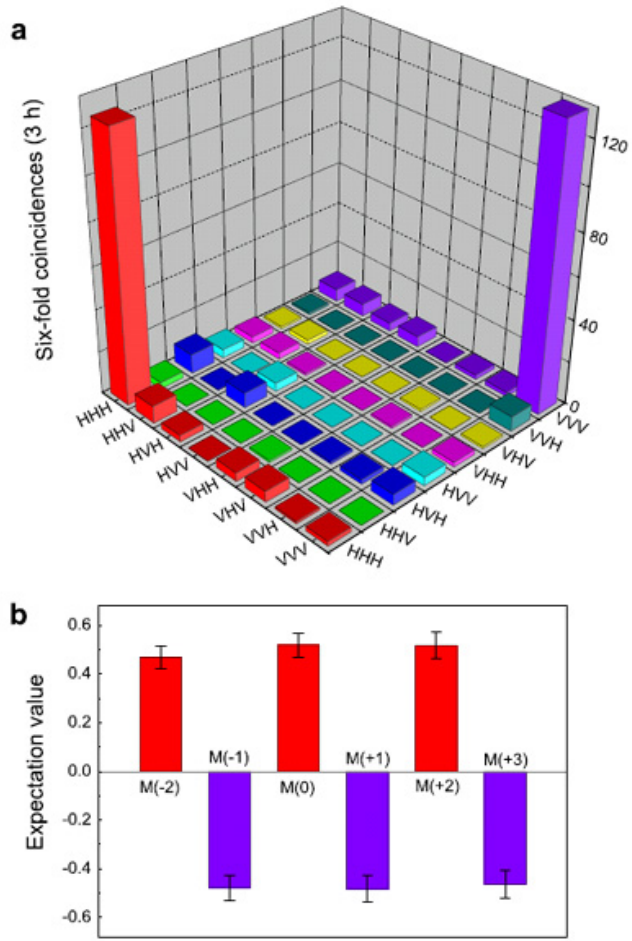

Figure 3

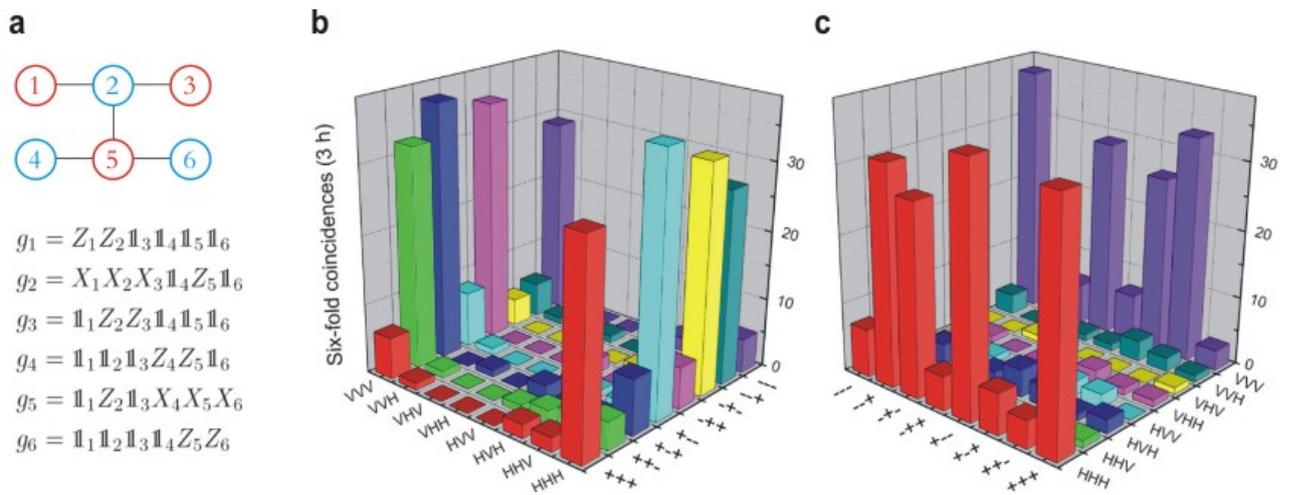

Figure 4

14